\documentclass[prd,twocolumn]{revtex4}
\usepackage{graphicx, epsfig}
\usepackage{color}
\usepackage{mathrsfs}
\usepackage{amsmath}

\newcommand{\be}{\begin{equation}}
\newcommand{\ee}{\end{equation}}
\newcommand{\bea}{\begin{eqnarray}}
\newcommand{\eea}{\end{eqnarray}}

\newcommand{\gapp}{\mathrel{\raise.3ex\hbox{$>$}\mkern-14mu
              \lower0.6ex\hbox{$\sim$}}}
\newcommand{\lapp}{\mathrel{\raise.3ex\hbox{$<$}\mkern-14mu
              \lower0.6ex\hbox{$\sim$}}}

\begin{document}

\title{The electroweak vacua, collider phenomenology and possible connection with dark energy}
\author{Eric Greenwood}
\author{Evan Halstead}
\author{Robert Poltis}
\author{Dejan Stojkovic}
\affiliation{HEPCOS, Department of Physics,
SUNY at Buffalo, Buffalo, NY 14260-1500}
\begin{abstract}
Higher dimensional non-renormalizable operators may modify the Standard Model Higgs potential in many interesting ways.
Here, we consider the appearance of a second vacuum which may play an important role in cosmology. For the certain range of parameters, the usual second order electroweak phase transition is followed by a first order phase transition that may drive the late time accelerated expansion of the universe. Such a potential contains kink-like solutions which in turn can play a crucial role in reconstructing the global shape of the potential in colliders, as we explicitly demonstrate.
\end{abstract}


\maketitle

\section{Introduction}

Today, observational data seems to indicate that our universe is going through a period of accelerated expansion. It remains a mystery, however, what is the driving force behind this acceleration. This problem is known as the dark energy problem.
Observationally, the equation of state for the universe is $w\approx-1$, which corresponds to a constant, or nearly constant, energy density. The minimal solution is the true vacuum energy density or cosmological constant. If this is indeed the case, this may represent the worst discrepancy between theory and observation. The value needed to explain the observed acceleration of the universe is $(10^{-3}$eV$)^4$. This value is some $124$ orders of magnitude smaller than the generic predicted value $(10^{19}$GeV$)^4$. Since our universe certainly does not have the generic value of the vacuum energy density, there is a serious question why is it so. This problem is known as the cosmological constant problem, and we would like to distinguish it from the dark energy problem. In particular, it is possible to construct models with the scalar field where the driving force behind acceleration is the scalar field vacuum energy density. Most of the models in the literature, including quintessence, k-essence and ghost-condensate, are such models \cite{Caldwell:1997ii,ArmendarizPicon:2000dh,ArkaniHamed:2003uy,Frieman:2008sn,Padmanabhan:2008wi,EspanaBonet:2008xd}. While these models can provide the mechanism for acceleration and resolve some issues connected with it, they can not solve the cosmological constant problem. In fact, no model dealing solely with the scalar field without addressing gravity can solve the cosmological constant problem. In the heart of the cosmological constant problem is the fact that the matter field Lagrangian is invariant under a shift by a constant. Such a shift does not change the equations of motion. However, gravity breaks that symmetry. Therefore, it is very unlikely that the cosmological problem will be solved without addressing gravity in a fundamental way \cite{Padmanabhan:2007xy}.

Here, we will adopt a common route. We will not try to solve the cosmological constant problem. We will simply propose a mechanism that can explain accelerated expansion of the universe using a scalar field, assuming that the cosmological constant problem is solved. The scalar field that is driving the acceleration does not have to be decoupled from the rest of the universe \cite{Stojkovic:2007dw,Chung:2007cn,Unnikrishnan:2008ki,Brax:2008as,Sami:2009dk}. It can, in fact, be the Standard Model Higgs field. The Higgs field is coupled to the other Standard Model particles which means that we can test the model in colliders.

The properties of the Standard Model Higgs potential are well known.  They depend on the Higgs mass and self-couplings. If we limit ourselves only to dimension-four renormalizable operators then the electroweak phase transition is second order \footnote{For small values of the Higgs mass \cite{Cline:2006ts} the usual electroweak phase transition can be first order. However, we keep here the standard assumption that the usual electroweak phase transition is the second order.}. However, it is clear that the Standard Model is only an effective low energy theory and there is no need to include only dimension-four operators. Inclusion of the higher dimensional non-renormalizable operators may introduce very interesting features. In particular, adding dimension-six and dimension-eight operators
can introduce two more symmetric minima. Then, for the certain range of parameters, the usual second order electroweak phase transition is followed by a first order phase transition. This subsequent phase transition may drive the late time accelerated expansion of the universe and yet leave the imprint in colliders. Here, we study such a scenario in detail.

\section{Model}

We consider the Standard Model Lagrangian for the Higgs field,  invariant under the electroweak transformations. We include the non-renormalizable dimension-six and dimension-eight operators
\be
  V(\Phi)=-\mu^2\Phi^{\dagger}\Phi+\lambda_1(\Phi^{\dagger}\Phi)^2-\lambda_2(\Phi^{\dagger}\Phi)^3+\lambda_3(\Phi^{\dagger}\Phi)^4+V_0
  \label{pot}
\ee
where $\Phi$ is the standard electroweak Higgs doublet. $V_0$ is a constant which is an overall shift of the potential.

At zero temperature the CP-even scalar state can be expanded in terms of its zero-temperature vacuum expectation value $v$ and the physical Higgs boson $H$:
\be \label{vev}
  \langle\phi\rangle=v,\hspace{3mm} \Phi=\frac{H+v}{\sqrt{2}}\equiv\frac{\phi}{\sqrt{2}}.
\ee
The potential as a function of $\phi$ is given by
\be
  V(\phi)=-\frac{\mu^2}{2}\phi^2+\frac{\lambda_1}{4}\phi^4-\frac{\lambda_2}{8}\phi^6+\frac{\lambda_3}{16}\phi^8+V_0.
  \label{Pot}
\ee
Substituting Eq.~(\ref{vev}) into Eq.~(\ref{Pot}) and reading off the coefficient in front of the quadratic terms in $H$, we can find the physical Higgs field mass $m_H$ as
\be \label{mh}
m_H= -\mu^2+3\lambda_1v^2+(7/2)\lambda_3v^6-(15/4)\lambda_2v^4 \, .
\ee
Defining
\bea
  \phi_1^2\equiv\frac{2}{\lambda_3\phi_2^2}\left(\lambda_1-\frac{1}{4}\frac{\lambda_2^2}{\lambda_3}\right),\nonumber\\
  \phi_2^2\equiv\frac{1}{2}\frac{\lambda_2}{\lambda_3}\left(1+\sqrt{3-\frac{8\lambda_3\lambda_1}{\lambda_2^2}}\right),\nonumber\\
  \varepsilon_0\equiv\frac{\mu^2}{2\phi_1^3\phi_2^3}-\frac{\lambda_3}{8\phi_1\phi_2}(\phi_1^2+\phi_2^2),
  \label{parameters}
\eea
we can then rewrite Eq.~(\ref{Pot}) in a more convenient way,
\be
  V(\phi)=-\varepsilon_0\phi_1^3\phi_2^3\phi^2+\frac{\lambda_3}{16}(\phi^2-\phi_1^2)^2(\phi^2-\phi_2^2)^2+V'_0
  \label{Veps}
\ee
where
\be
  V'_0\equiv V_0-\frac{\lambda_3}{16}\phi_1^4\phi_2^4.
\ee
Here we note that we have restriction on $\lambda_1$. From Eq.~(\ref{parameters}), requiring that both $\phi_1,\phi_2$ be real, we have the requirement that
\be
  \frac{1}{4}\frac{\lambda_2^2}{\lambda_3}<\lambda_1<\frac{3}{8}\frac{\lambda_2^2}{\lambda_3}.
  \label{restriction}
\ee

The role of the parameter $\varepsilon_0$ is to introduce a controlled fine tuning of $V(\phi)$. If $\varepsilon_0=0$, the potential in Eq.~(\ref{Veps}) has two degenerate minima at $\phi=\phi_1$ and $\phi=\phi_2$. If $\varepsilon_0 \neq 0$, the difference between the energy densities of the two vacua is
\be
  \delta V=\varepsilon_0\phi_1^3\phi_2^3(\phi_2^2-\phi_1^2).
  \label{DV}
\ee
For the sake of definiteness, we take that $\phi_2 > \phi_1$.

$V_0$ has been added to the potential to specify the false vacuum energy density. We require that  $V (\phi_1) \approx(10^{-3}$eV$)^4$. This choice represents a vacuum energy density that is sufficient to drive the accelerated expansion of the universe. We do not explain the appearance of such small number. The solution to the true cosmological constant problem may explain it. For example, an interesting numerology $10^{-3}$eV$ \approx (TeV/M_{Pl})TeV$ hints toward a gravitational origin
of this small number. In particular, operators suppressed by powers of $M_{Pl}$ might be responsible for it.

$V(\phi)$ in Eq.~(\ref{Veps}) is the zero-temperature potential.
In order to study the sequence of phase transitions we need to calculate the finite-temperature effective potential.
Finite temperature effects are approximated by adding a thermal mass to the potential. The potential is then written as $V(\phi,T)=cT^2\phi^2/2+V(\phi,0)$, where $c$ is generated by the quadratic terms that acquire a $\phi$-dependent mass in the high-$T$ expansion of the one-loop thermal potential. Note that there are also terms which are proportional to $T^2\phi^4$, however these terms only lead to small corrections to the potential \cite{Grojean:2004xa}.

If the mass of a certain species of particles is greater than the temperature of the plasma, the thermal corrections due to this species decouple exponentially. Therefore, strictly speaking, one has to multiply the contribution from each of the species by the step function $\Theta (T-m)$ \cite{Comelli:1996vm}. While this effect may modify the fine details of the phase transition the general qualitative features will remain unchanged.

The effective potential can then be written as
\be \label{veff}
  V_{eff}(\phi,T)\equiv-\varepsilon (T)\phi_1^3\phi_2^3\phi^2+\frac{\lambda_3}{16}(\phi^2-\phi_1^2)^2(\phi^2-\phi_2^2)^2+V'_0.
\ee
where
\be \label{et}
  \varepsilon (T)=\varepsilon_0-\frac{cT^2}{\phi_1^3\phi_2^3}.
\ee
Following the procedure in \cite{Viswanathan} and \cite{Vilenkin}, the constant $c$ is given by
\be \label{c}
  c=\frac{1}{16}\left(3g^2+g'^2+4y_t^2+\frac{1}{32}\lambda_1\right)
\ee
$g$ and $g'$ are the $SU(2)_{L}$ and $U(1)_{Y}$ gauge couplings, and $y_t$ is the top Yukawa coupling. All temperature dependence is in $\varepsilon (T)$.

\begin{figure}[ht]
\includegraphics{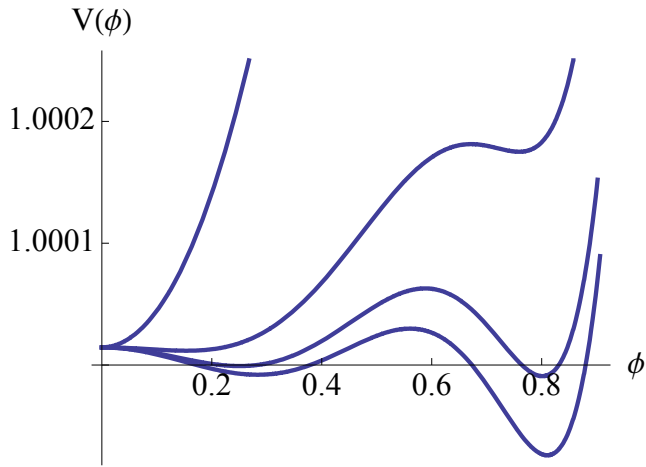}
\caption{The temperature dependence of the $\phi^8$ Higgs field potential. As the temperature decreases the minimum at $\phi=0$ becomes the maximum and the field starts rolling down. Simultaneously two new minima appear. The standard electroweak (second order) phase transition is over  when the Higgs field ends up in the first vacuum at $\phi=\phi_1$. Then the second minimum at $\phi=\phi_2$ descends and becomes the true vacuum. This drives the first order phase transition that may explain the late time accelerated expansion of the universe. The picture is symmetric with respect to the vertical axis. The values used in plot are: $\phi_1 = 0.246$TeV, $\phi_2 = 0.8$TeV, $\lambda_3 = 0.154$TeV$^{-4}$ and $\varepsilon_0 = 0.015$TeV$^{-4}$.}
\label{T-dep}
\end{figure}

Now, we can study that change of the shape of the potential as the universe cools down (as shown in Fig.~\ref{T-dep}). At very high temperatures, before the electroweak symmetry breaking, the potential has a characteristic "U" shape with a single minimum at $\phi=0$.  The whole potential is symmetric and we can consider only $\phi >0$ semi-plane.
The vacuum energy density of the Higgs field before the electroweak phase transition is generically of the order of the characteristic energy scale of the phase transition, i.e. $\sim 100$GeV.
However, notice that the Higgs field is sitting there only before the electroweak phase transition where the temperature of the universe is high and the universe is radiation dominated.
The Higgs field has zero expectation value and the electroweak symmetry is not broken. As $T$ falls, the minimum at  $\phi=0$ becomes a maximum, while simultaneously two new minima appear. The Higgs field then rolls down the potential to the first minimum at $\phi=\phi_1$. There, the Higgs field has a non-zero expectation value and the electroweak symmetry is broken. This is the standard electroweak phase transition which is of the second order. Thus, we do not change the standard picture of the early universe. We currently live in $\phi=\phi_1$ vacuum, where the vacuum expatiation value of the Higgs field is $\phi_1 = 246$GeV.

However, the temperature dependent evolution of the Higgs field potential does not stop there. The other minimum at $\phi=\phi_2$ keeps descending as the temperature in the universe drops. Eventually, below some critical temperate, the minimum at $\phi=\phi_2$ becomes the true global minimum, while the minimum at $\phi=\phi_1$, that we live in, becomes a false minimum.

In order to justify an effective field theory description of the non-renormalizable operators, it is important that the coefficients with negative mass dimension in the potential are considerably less than unity (in units of TeV, assuming that the new physics comes at the TeV scale), as is the case for the values in the plot in Fig.~\ref{T-dep}. The second requirement is that the scalar field vacuum expectation values are lower than the TeV scale. In the plot in Fig.~\ref{T-dep}, the value of the second vacuum expectation value is $\phi_2 = 0.8$TeV which is getting close to the new physics scale. Note that the values chosen for the plot are just for the purpose of illustration, more detailed analysis of the possible values of the parameters in the potential will be done in Section~\ref{values}.

The critical temperature is given by $\varepsilon (T)=0$. From Eq. {\ref{et}) we get
\be \label{tc}
  T_c^2=\frac{\varepsilon_0\phi_1^3\phi_2^3}{c}.
\ee
At $T=T_c$ we have $\delta V=0$ and the heights of the two minima are equal. For $T<T_c$, the second minimum at $\phi_2$ becomes the true minimum with the energy density difference between the vacua given by Eq.~(\ref{DV}).

At high temperatures there is also an overall correction to the potential that is proportional to $N T^4$, where $N$ is the number of relativistic particle species in the plasma. This contribution could modify the critical temperature if there are different numbers of relativistic species in the two adjacent vacua \cite{Quiros:1999jp,Carena:2004ha,Craig:2006kx}. We examine this effect in Appendix~\ref{ct}.

The existence of the lower minimum than the one we currently live in indicates that our universe will eventually tunnel into the true vacuum. This phase transition may drive the late time accelerated expansion of the universe. The tunneling rate depends on the energy difference between the vacua which in turn depends on the parameter $\varepsilon_0$.

For completeness, we note that one could set $\phi_1 > \phi_2$, with $\phi_1$ identified with the standard electroweak vacuum. In this case, the electroweak phase transition will be more complicated than usually thought, since it would be composed of two subsequent phase transitions before the Higgs field settles down into today's vacuum $\phi_1$. This interesting possibility is outside of the scope of this paper.

\section{Tunneling Rate}
\label{TR}

An important question is how long will the universe exist in the false vacuum state $\phi=\phi_1$? The transition from the false to the true vacuum occurs from nucleation of bubbles of true vacuum inside the false vacuum. The transition probability per unit space-time volume, using the semi-classical approximation, is given by
\be
  \Gamma=Ae^{-S_E}
  \label{tunnel}
\ee
where $S_E$ is the Euclidean action of the bounce solution and $A$ is a dimensionful constant, which depends on the loop corrections to the potential Eq.~(\ref{Pot}). However, here we are only interested in order of magnitude transition rate, thus we will ignore these corrections. To calculate $S_E$, we follow the method developed by Coleman \cite{Coleman}.

The one-dimensional Euclidean action per unit volume for the tunneling is, to zeroth order in $\varepsilon_0$
\bea
  S_1&=&\int_{\phi_1}^{\phi_2}d\phi'\sqrt{2V(\phi')}\nonumber\\
        &\approx&\frac{2}{15}\sqrt{\frac{\lambda_3}{8}}(\phi_2-\phi_1)^3(\phi_1^2+3\phi_1\phi_2+\phi_2^2)
\eea
where we are again assuming that $\phi_2 > \phi_1$. In the zero-temperature limit and thin wall approximation, the radius the critical bubble is then
\be
  R_0=\frac{3S_1}{\delta V}=\frac{3}{5}\sqrt{\frac{\lambda_3}{8}}\frac{(\phi_2-\phi_1)^2(\phi_1^2+3\phi_1\phi_2+\phi_2^2)}{\varepsilon_0(\phi_1+\phi_2)\phi_1^3\phi_2^3}.
  \label{radius}
\ee
For an $O(4)$ symmetric bubble, the Euclidean action is then given by
\bea
\label{SE}
  S_E&=&-\frac{1}{2}\delta V\pi^2R_0^4+2\pi^2R_0^3S_1\nonumber\\
       &=&\frac{8\lambda_3^2}{120000}\frac{\pi^2(\phi_2-\phi_1)^9(\phi_1^2+3\phi_1\phi_2+\phi_2^2)^4}{\varepsilon_0^3(\phi_1+\phi_2)^3\phi_1^9\phi_2^9}.
\eea
At zero temperature, Eq.~(\ref{tunnel}) gives the decay rate per unit volume per unit time. In order for our observable universe, whose four-volume is of the order of $t^4_{Hubble}$, to remain in the false vacuum, one must require that $\Gamma t^4_{Hubble}\leq1$. Taking $t_{Hubble}\sim10^{10}$years, we find that sufficient stability for the false vacuum is obtained for $S_{E}>400$. With the generic value $\lambda_1\sim1$, we see that vacuum stability requires only $\varepsilon_0 \leq 0.012$TeV$^{-4}$. This is also enough to make the thin wall approximation valid.

To ensure that the above analysis remains true for the early universe, we calculate the temperature dependent decay rate in the high temperature limit. At finite temperatures the $O(4)$ symmetric bounce is approximately the periodic in time $O(3)$ symmetric solution. In this solution, the period is $1/T$ (see \cite{Linde}). The decay exponent, i.e. the Euclidean action, now has the form
\be
  S_E=\frac{S_3(\phi,T)}{T}
\ee
where $S_3$ is the three-dimensional action of an $O(3)$ bubble. The new radius of the critical bubble is now
\bea
  R(T)&=&\frac{2S_1}{\delta V(T)}\nonumber\\
      &=&\frac{4}{15}\sqrt{\frac{\lambda_3}{8}}\frac{(\phi_2-\phi_1)^2(\phi_1^2+3\phi_1\phi_2+\phi_2^2)}{\varepsilon(T)(\phi_1+\phi_2)\phi_1^3\phi_2^3}.
\eea
The three-dimensional action of the $O(3)$ bounce solution is then given by
\bea
  S_3(T)&=&-\frac{4}{3}\pi R(T)^3\delta V(T)+4\pi R(T)^2S_1\nonumber\\
         &=&B\frac{\pi(\phi_2-\phi_1)^7(\phi_1^2+3\phi_1\phi_2+\phi_2^2)^3}{\varepsilon(T)^2(\phi_1+\phi_2)^2\phi_1^6\phi_2^6}
         \label{S3}
\eea
where
\be
  B\equiv\frac{128\lambda_3^{3/2}}{10125\sqrt{512}}.
  \label{B}
\ee
Therefore the temperature dependent decay rate is given by Eq.~(\ref{tunnel}), Eq.~(\ref{S3}) and Eq.~(\ref{B}) as
\be \label{tddr}
\Gamma \sim \exp \left( -\frac{{\rm const}}{T\varepsilon (T)^2}  \right)
\ee
where $\varepsilon (T)$ is given in Eq.~(\ref{et}).
We can see here some important properties. First, at high temperature, $T>>T_c$, when $\phi=\phi_1$ is the lowest energy state, the transition rate is large. Thus most of the universe ends up in that minimum. At $T=T_c$, we have that $\varepsilon (T)=0$, thus the transitions between the vacua are suppressed.
The high temperature approximation is valid, at least formally, at
the temperatures slightly below $T_c$. It is there that the
decay rate in Eq.~(\ref{tddr}) is maximal and we need to correct the
zero temperature estimate for $\varepsilon_0$. Fortunately, a slight
correction $\varepsilon_0 \sim 0.005$TeV$^{-4}$ makes the decay rate safely small. As shown earlier, for $T \ll T_c$, $\phi=\phi_1$  is a false minimum, but the transition rate to the true vacuum at $\phi=\phi_2$ is suppressed by the bare value of $\varepsilon_0$. We saw that
$\varepsilon_0 \sim 0.01$TeV$^{-4}$ makes the transition time larger than the current
Hubble time. In order to incorporate a somewhat stronger
constraint for high temperatures, we require  $\varepsilon_0 \sim 0.005$TeV$^{-4}$.

We have to make sure that energetic processes
in our universe (e.g. cosmic ray collisions) were not able to initiate the formation of a true vacuum bubble, which would in turn encompass most of the visible universe by now.
Fortunately, it is not a simple thing to create a vacuum
bubble in a high energy collision. This requires not just
sufficient energy but a coherent superposition of a large
number of high energy quanta over a volume large compared
to the characteristic energy. Such processes require high densities and high temperatures, not only high energies.
The height of the barrier between the false and true vacua is
\be
  V_{max}=\frac{\lambda_3}{256}(\phi_2^2-\phi_1^2)^4.
  \label{Vmax}
\ee
This is approximately $(0.1$TeV)$^4$ for the values $\phi_1\sim 0.246$TeV, $\phi_2\sim 0.8$TeV and $\lambda_3\sim 0.154$TeV$^{-4}$. Such temperatures are unlikely to soon be achieved in colliders, and are probably not achieved over large enough volume even in the highest energy cosmic ray collisions.

The critical bubble radius, Eq.~(\ref{radius}), is $107$TeV$^{-1}$ for our parameters ($\varepsilon_0\sim0.098$TeV$^{-4}$, $\phi_1\sim0.246$TeV, $\phi_2\sim 0.8$TeV and $\lambda_3\sim0.154$TeV$^{-4}$). Given Eq.~(\ref{Vmax}), this suggests that we need approximately
\be
  N_{quanta}\simeq\left(\frac{4\pi R_0^3}{3}V_{max}\right)V_{max}^{-1/4}>10^9(1/\varepsilon_0)^3.
\ee
individual excitations coherently superimposed. Moreover, $N_{quanta}$  grows very fast, as $\varepsilon_0^{-3}$, making it extremely difficult to create a critical bubble in a high energy collision.

In the case of tunneling from dS to AdS vacuum, there is an additional suppression due the different asymptotics of these space-times.  The size of the critical bubble in AdS ends up being larger than one might generically expect \cite{Johnson:2008kc}, which further suppresses the transition.

\section{Future of the universe}

In this context we can address the question of the future of our universe. Phase transitions give quite different predictions for the future of our universe than the true cosmological constant. In the case of the true cosmological constant, accelerated expansion never stops. Acceleration will slowly drive most of  today's visible universe out of the cosmological horizon. In a distant enough future, the whole visible universe will be a gravitationally bound system consisting of only Milky Way and Andromeda galaxies \cite{Adams:1996xe,Krauss:2007nt}.

However, in the context of phase transitions we have several different possibilities. First, if the difference between the two vacua is very small (i.e the parameter $\varepsilon_0 \sim 0$), the phase transition will never be completed.
The bubble nucleation rate will be very slow and they will never percolate, since the background is expanding with acceleration.
This scenario is similar to the true cosmological constant, except for the possibility to have a few bubbles here and there locally.

The more interesting case is when the parameter $\varepsilon_0$ takes much less fine tuned values, for example those which allow for the phase transition to be completed.
Phase transitions are violent events and many things will be different in the new vacuum after the completion of the phase transition. The true vacuum at $\phi=\phi_2$ clearly has a different Higgs field vacuum expectation value that the vacuum we currently live in. This means that most of the Standard Model particles will have different masses. In such a universe it is very difficult to imagine life similar to ours due to the well known anthropic reasons.

Finally, for the perhaps most generic value of $\varepsilon_0 \sim 0.01$TeV$^{-4}$, the phase transition is just about to happen, in cosmological terms. Since the characteristic scale in the Higgs potential is of the order of $100$GeV, the requirement that the false vacuum at  $\phi=\phi_1$ is shifted up from zero by a tiny amount of $10^{-3}$eV directly implies that the true vacuum at $\phi=\phi_2$ is deeply AdS, i.e. has a negative vacuum energy density \cite{Cvetic:1993xe,Ansoldi:2007qu,Metaxas:2008dn,Lee:2008hz}. The transition from the false to the true vacuum will be described by the Coleman-De Luccia instanton \cite{Coleman:1980aw}. According to \cite{Abbott:1985kr}, any initial instabilities in the AdS space will quickly grow and cause the collapse of the whole universe into a black hole. In such a scenario it is difficult to imagine any life at all.

\section{Constraining the values of parameters in the potential}
\label{values}

The original potential has four parameters: $\mu$, $\lambda_1$, $\lambda_2$ and $\lambda_3$. It would be interesting to find the possible values that do not lead to dangerous exotic vacua.
The main constraint comes from Eq.~(\ref{SE}), where we require that $S_E \geq 400$. The requirement that today we live in the standard electroweak vacuum is that $\phi_1=246$GeV. From Eq.~(\ref{parameters}), $\phi_1$ can be expressed as $\phi_1 = \phi_1 (\lambda_1, \lambda_2, \lambda_3)$, which puts one constraint on the possible values of parameters. From this constraint we can express $\lambda_3$ in terms of $\lambda_1$ and $\lambda_2$. In order to make a useful plot of $S_E$ in terms of the two parameters, we need to fix one more parameter. We do that for $\mu$ by setting the value of the Higgs mass ($M_{\rm Higgs} = \sqrt{2} \mu$). Since the exact value of the Higgs mass is not known, we will make two plots, one assuming that the Higgs mass takes its lowest (experimentally) allowed value of $114$GeV (Fig.~ \ref{rp1}) and the other one assuming that the Higgs mass takes the value of $200$GeV (Fig.~ \ref{rp2}).

If we set the value of the Higgs mass to $114$GeV, from Fig.~ \ref{rp1} we see that possible ranges of the allowed values for the parameter $\lambda_1$ and $\lambda_2$ are $0<\lambda_1<0.2$, and  $0<\lambda_2<1$TeV$^{-2}$.
Incorporating these limits into the constraint equation $\phi_1 = \phi_1 (\lambda_1, \lambda_2, \lambda_3)$, we get that a possible range of the allowed values for the parameter $\lambda_3$ is given by $0<\lambda_3<10 TeV^{-4}$, as shown in Fig.~ \ref{l31}.

\begin{figure}[ht]
\includegraphics[width=2.8in]{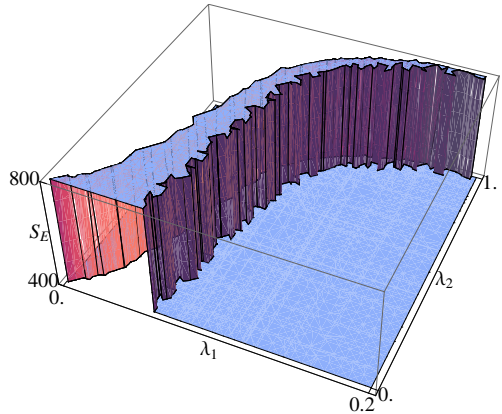}
\caption{A scan over the allowed range of parameters $\lambda_1$ and $\lambda_2$ in the original potential (\ref{Pot}) which do not lead to the phenomenologically excluded new vacua. The constraint comes from the requirement that the Euclidean action $S_E \geq 400$. For this plot, we set the value of the Higgs mass to $114$GeV.}
\label{rp1}
\end{figure}

\begin{figure}[ht]
\includegraphics[width=2.8in]{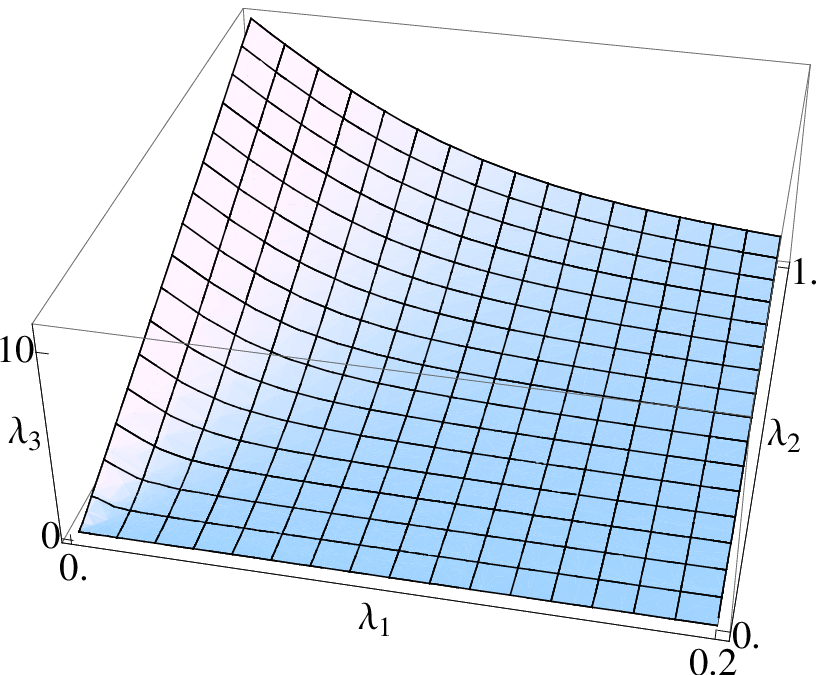}
\caption{The allowed range of the parameter $\lambda_3$ in the original potential (\ref{Pot}) which does not lead to the phenomenologically excluded new vacua. We use Eq.~(\ref{parameters}) and the requirement that $\phi_1=246$GeV to express $\lambda_3$ in terms of $\lambda_1$ and $\lambda_2$. For this plot, we set the value of the Higgs mass to $114$GeV.}
\label{l31}
\end{figure}

\begin{figure}[ht]
\includegraphics[width=2.8in]{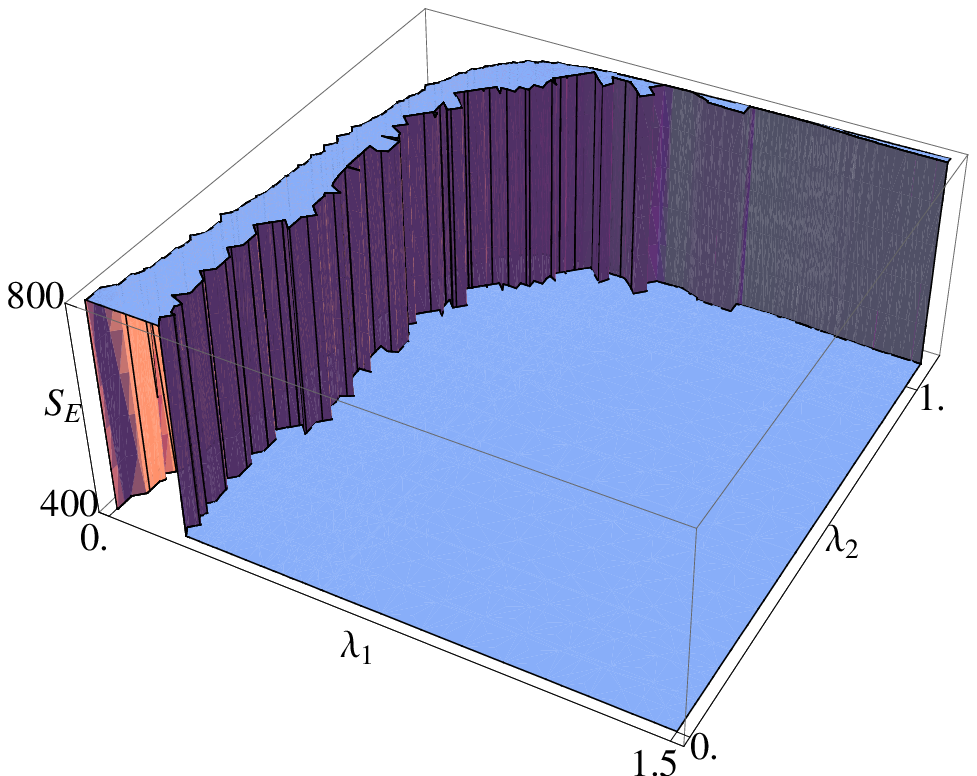}
\caption{A scan over the allowed range of parameters $\lambda_1$ and $\lambda_2$ in the original potential (\ref{Pot}) which do not lead to the phenomenologically excluded new vacua. The constraint comes from the requirement that the Euclidean action $S_E \geq 400$. For this plot, we set the value of the Higgs mass to $200$GeV.}
\label{rp2}
\end{figure}

\begin{figure}[ht]
\includegraphics[width=2.8in]{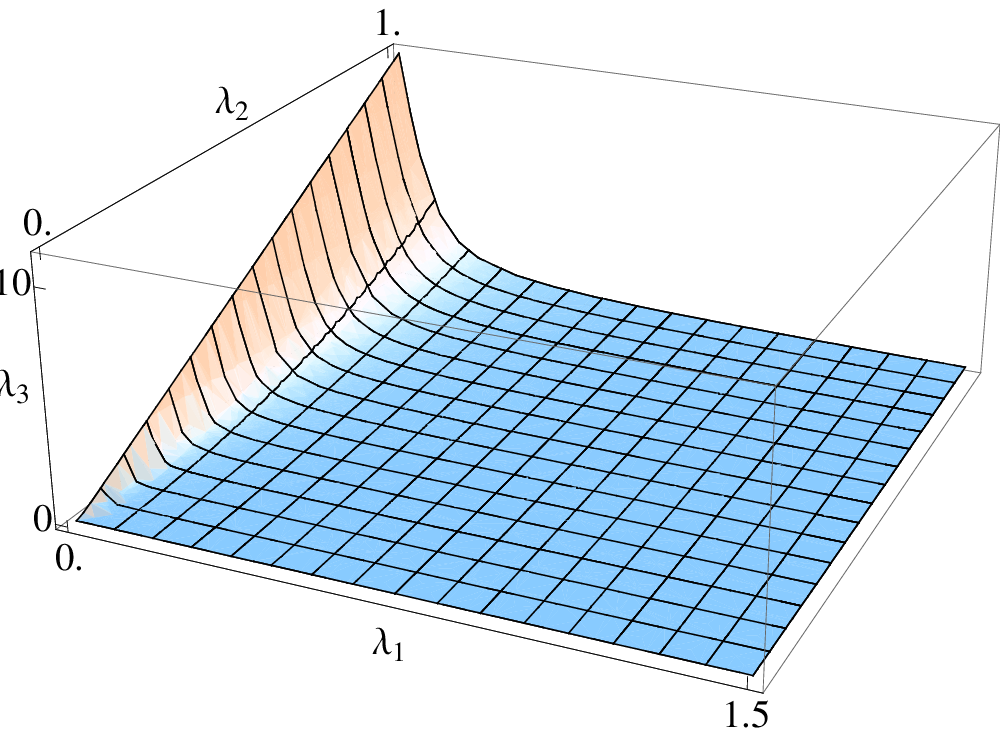}
\caption{The allowed range of the parameter $\lambda_3$ in the original potential (\ref{Pot}) which does not lead to the phenomenologically excluded new vacua. We use Eq.~(\ref{parameters}) and the requirement that $\phi_1=246$GeV to express $\lambda_3$ in terms of $\lambda_1$ and $\lambda_2$. For this plot, we set the value of the Higgs mass to $200$GeV.}
\label{l32}
\end{figure}

If we set the value of the Higgs mass to $200$GeV, from Fig.~ \ref{rp2} we see that possible ranges of the allowed values for the parameter $\lambda_1$ and $\lambda_2$ are $1<\lambda_1<1.5$, and  $0<\lambda_2<1$TeV$^{-2}$.
Incorporating these limits into the constraint equation $\phi_1 = \phi_1 (\lambda_1, \lambda_2, \lambda_3)$, we get that a possible range of the allowed values for the parameter $\lambda_3$ is given by $0<\lambda_3<10$TeV$^{-4}$, as shown in Fig.~ \ref{l32}.
As mentioned before, the effective field theory description of the non-renormalizable operators is valid only for the values of  $\lambda_2, \lambda_3 <1$ (in units of TeV, assuming that the new physics comes at the TeV scale).

Having the allowed range of parameters in the potential we can estimate the range of values that the second vacuum $\phi_2$ can take. From (\ref{parameters}) we can see that $\phi_2$ can take
values from $300$GeV to well above a TeV. Again, the effective field theory description is valid only for the values of $\phi_2$ smaller than TeV.

\section{Reconstructing the potential in colliders}

The main feature that distinguishes this type of models from other scalar field models (say quintessence) is the explicit connection with particle physics and the possibility of reconstructing the potential in colliders. With the inception of the LHC, we hope to be able to achieve this there.

In this model, we explicitly used the $\phi^8$ potential. How do we probe the global structure of the potential? If we excite the field locally, only around one vacuum, then we can only probe the local structure of the potential. In order to probe the global shape of the potential we need to excite a solution that extrapolates between the vacua. These are non-perturbative kink-like solutions.
In \cite{Vachaspati}, the author constructed a recipe for the reconstruction of the potential when kink-like solutions are present in the theory. The base of the recipe is the inverse scattering method.
In the direct scattering methods we start from the known potential and calculate the eigenfrequencies, i.e. the energies of the scattered particles. In the inverse scattering method, we start from the known eigenfrequencies (acquired presumably in the scattering experiments) and calculate the shape of the potential which is the cause of the scattering.

In our context, kink-like solutions which connect the two vacua correspond to the bubbles of the true vacuum. Inside the bubble we have the true vacuum while outside is the false vacuum. The true vacuum is energetically favored so the volume term contributes to the pressure directed outward. However, the surface tension of the bubble tends to contract the bubble. The critical bubble is the one which is large enough so that these two forces are balanced.  We showed above that the production of the critical bubble of the true vacuum which is capable of expanding is very suppressed. However, production of a subcritical bubble which will collapse under its own tension should not be severely suppressed. By studying production and decay of these subcritical bubbles we can learn a lot about the global shape of the potential.

Bubbles of true vacuum are soliton-like solutions, and in zeroth order the production cross section should be just the geometrical cross section, i.e. $\pi R_{\rm bubble}^2$ where $R_{\rm bubble}$ is the geometrical radius of the bubble. The situation is somewhat similar to mini-black hole production in high energy collisions. Black holes are gravitational solitons and their production cross section was shown to be just the geometrical cross section, i.e. $\pi R_{BH}^2$ \cite{Eardley:2002re}. However, there is one significant difference. While in the of case mini black hole production, according to the hoop conjecture, there are no phase space suppression factors, in the case of the bubble the suppression factor must be present. A bubble is a coherent superposition of a certain number of the scalar field quanta, say $N$. Therefore the suppression factor would likely go as $e^{-{\rm entropy}}$ which is roughly $e^{-\ln(N!)} \sim e^{-N}$.  For a subcritical bubble where $N$ is a few, the production may be possible \cite{Stojkovic:2001qi}. For a critical bubble with large $N$, the production is highly suppressed. Note also that the threshold for the bubble production would be $Nm$, where $m$ is the mass of the scalar field quanta.

If the bubble of the true vacuum is produced, it is unlikely that it will be produced in its ground state. Instead, we expect that it will be produced in a highly excited state. Then the decay of such a bubble will give off the eigenvalues of the potential.
Since bubbles are coherent states of a certain number of the scalar field quanta, they would dominantly decay into those scalar field quanta (in this case the Higgs field). The eigenvalues of the potential would come from the (reconstructed) energy distribution of emitted scalar field quanta. Due to the spherically symmetric configuration of the bubble, one of the main signatures of the bubble production would be a spherically symmetric distribution of emitted scalar field quanta. For the light Higgs, the main decay channel will be through b-quarks, which further decay to a c-quark, a lepton (which may serve as a trigger) and neutrino. In that case, b-quarks will be produced copiously. Because of the high multiplicity, b-quarks may not be very energetic. The main question is then whether this signature can be distinguished from the QCD background. Two things are important in this context: first, high degree of spherical symmetry, and second, many jet events. For example, for five (and likely more) jet events the trigger threshold may be lowered to about $50$GeV, in which case the QCD background can be kept under control. The heavier Higgs is much easier to analyze. For the heavy Higgs, decay channels including $W^\pm$ and $Z$ gauge bosons becomes significant. Once the $ZZ$ branching ratio become significant, the experimental signature is much more distinct. The other question is the possibility of the reconstruction of the original energy of the bubble through the decay products. Situations containing neutrinos in the final stage are specially inconvenient. However, based on earlier experience, say with the top quark, we know that something like that is possible. Analyzing other decay products, one may identify the exact channel of decay and thus estimate the energy taken by neutrinos.

If a scalar field theory is written in the standard form as
\be
L = \frac{1}{2} \partial_\mu \phi \partial^\mu \phi - V(\phi)
\ee
then the equations of motion can be written in the Schrodinger-like form
\be \label{eom}
\left[ - \frac{d^2}{dx^2} + \frac{d^2  V(\phi_0(x))}{dx^2}  \right] \psi_n (x) = \omega_n^2 \psi_n(x).
\ee
For simplicity, the field $\phi (x)$ is a function of only one coordinate. $\phi_0 (x)$ is the (unknown) profile function of the
kink solution. The task is to determine the potential $V(\phi)$ knowing the eigenvalues $\omega_n$. Though the answer is not unique, additional theoretical input (e.g. total energy of the bubble and some perturbative interactions) can possibly reduce degeneracies.
The zero mode $\psi_0 = d\phi_0/dx$ which corresponds to the eigenvalue $\omega_0 =0$ plays a special role. In theories where kink solutions are present the Bogomolnyi equation directly relates the zero mode to the shape of the potential
\be
\psi_0 = \frac{d \phi_0(x)}{dx} = \pm \sqrt{2V(\phi_0)} .
\ee
Therefore
\be
V(\phi_0) = \frac{1}{2} \psi_0(x)|_{x(\phi_0)} ,
\ee
where $x(\phi_0)$ is obtained by inverting $\phi_0(x)$ which is in turn obtained by integrating the relation $\psi_0 = d\phi_0/dx$. Thus, by finding the zero mode solution to the Eq.~(\ref{eom})
we can reconstruct the shape of the potential. The problem is that, in most of the cases, the zero mode is coupled to all of the higher modes, and we need to solve a coupled set of differential equations.
However, as shown in \cite{Vachaspati}, two of the most important theories containing kink solutions, i.e. sine-Gordon and  $\phi^4$ theory, have respectively one and two eigenfrequencies. In these cases we can solve the equations analytically. In more complicated cases, it is possible that the equations need to be solved numerically.

We now come back to our $\phi^8$ potential. Since the original potential has four parameters $(\mu, \lambda_1,\lambda_2,\lambda_3)$, we can infer that in this case there are four eigenfrequencies: $\kappa_1$, $\kappa_2$, $\kappa_3$, and $\kappa_4$. Following the procedure in \cite{Vachaspati}, in the special with a high degree of symmetry between the eigenvalues, one can reconstruct the potential (for details see Appendix A)
\be \label{rp}
  V(\phi_0)=\frac{\alpha^2}{2}\left(1-\frac{9\phi_0^2}{4\alpha^2}\right)^4
\ee
where $\alpha$ is a normalization constant. Here we recognize the $\phi^8$ potential. In order to get a desired potential with two pairs of non-equivalent minima we need to relax the conditions on the eigenvalues. In this more generic case, the potential can not be obtained in an analytic form since the differential equations must be solved numerically. The result of the calculations presented in the Appendix A is shown in Fig.~\ref{Recon}.
\begin{figure}[htbp]
\includegraphics{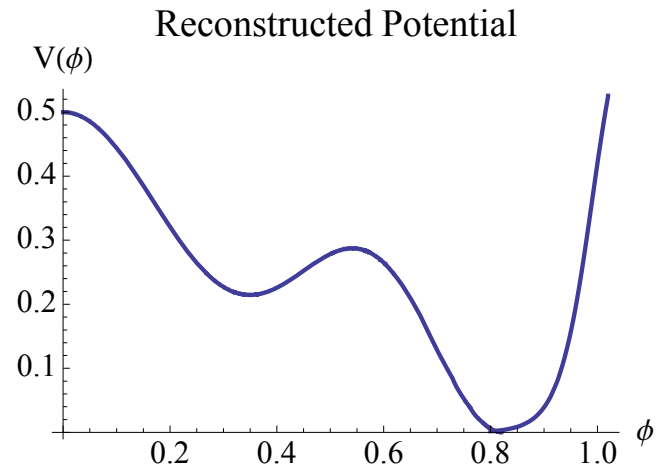}
\caption{Reconstruction of the $\phi^8$ potential using the inverse scattering method. The specific values for this plot are $\nu=1$, $\lambda=-4$, $\kappa_2^2-\kappa_3^2=5$ and $\kappa_3^2-\kappa_4^2=\kappa_3^2=3$. The plot is symmetric with respect to vertical axis.}
\label{Recon}
\end{figure}
By comparing the potential in Eq.~(\ref{Veps}) with the reconstructed one in Fig.~\ref{Recon}, we can infer the values of the parameters in our original potential. In particular, we can obtain the value of the fine tuning parameter $\varepsilon_0$. If the solution to the true cosmological constant implies that the true vacuum energy density vanishes, then the parameter $\varepsilon_0$ must give for the energy density difference between the vacua in Eq.~(\ref{DV}) the value of
$\delta V \approx (10^{-3}$eV$)^4$.

\section{Conclusion}

The electroweak Standard Model is believed to be only a low energy effective theory. For this reason we may be allowed to introduce higher dimensional non-renormalizable operators. We expect new physics to kick in close to a TeV scale. Here we studied how
these higher dimensional non-renormalizable operators may modify the Standard Model Higgs potential in a manner  interesting for cosmology. We considered the appearance of a second vacuum in the Higgs field potential. We calculated finite temperature corrections  and showed that the usual second order electroweak phase transition is followed by a first order phase transition that may drive the late time accelerated expansion of the universe. Such a potential contains kink-like solutions which in turn can play a crucial role in reconstructing the global shape of the potential in colliders. We explicitly demonstrated it using the inverse scattering method adopted to studies of theories where kink-like solutions are present.

We addressed the future of our universe in this context which is quite different from the future dictated by the true cosmological constant.

Since our model does not address gravity in a fundamental way, it does not solve the cosmological constant problem. It simply addresses the dark energy problem as all of the other scalar field models do --- postulates the potential consistent with a given equation of state, which in this case is $w=-1$. The advantage here is that we do not need to postulate the existence of a new scalar field completely decoupled from the rest of the universe. As an additional bonus in this model, the potential can be in principle reconstructed in colliders.

While we do not explain the appearance of a small number $10^{-3}$ an interesting numerology $10^{-3}$eV$ \approx ($TeV/M$_{Pl}$)TeV hints toward a gravitational origin of this small number. Operators suppressed by powers of M$_{Pl}$ might be responsible for it.

\begin{acknowledgments} The authors are grateful to U. Baur and A. Kharchilava for very useful conversations. DS acknowledges the financial support from NSF.
\end{acknowledgments}

\appendix

\section{Reconstruction of Potential}

We illustrate here the inverse scattering problem, i.e. reconstruction of the potential from its eigenvalues.
Knowing that the $\phi^4$ potential has two eigenvalues, we expect that he $\phi^8$ potential has four. We take the four eigenvalues of the bound-states to be $\kappa_1$, $\kappa_2$, $\kappa_3$, and $\kappa_4$. We will first consider the special case where the solution can be found analytically, then we will do numerical analysis of a more generic case.

The translational mode is always the lowest, so we can write $\kappa_4=0$, since we are using the notation $\kappa_i>\kappa_{i+1}$. Following \cite{Vachaspati}, we find the potential containing $n$ of the highest bound states:
\be
  U_n(x)=f_n^2+f'_n+\kappa_n^2
  \label{U}
\ee
where the function $f_n(x)$ satisfies
\be
  f'_n-f_n^2+U_{n-1}=\kappa_n^2.
  \label{f_n}
\ee
Defining $f_n(x)\equiv-w'_n/w_n$ we can rewrite Eq.~(\ref{f_n}) as
\be
  -w''_n+U_{n-1}w_n=\kappa_n^2w_n.
\ee
This equation will have two linearly independent solutions; however, if we require that $U_n$ is even under parity transformations then we must also require $w_n(-x)=w_n(x)$. This condition then eliminates one of the linearly independent solutions.

Therefore to find $U_1$ we consider
\be
  U_1=f_1^2+f'_1+\kappa_1^2
  \label{U1}
\ee
and solve
\be
  -w_1''+U_0w_1=\kappa_1^2w_1.
  \label{w1}
\ee
Now if we take $U_0=\kappa_0^2$ and define $\nu^2\equiv\kappa_0^2-\kappa_1^2$ we can then write Eq.~(\ref{w1}) as
\be
  -w''_1+\nu^2w_1=0.
\ee
The solution to this is $w_1=\cosh(\nu x)$, therefore
\be
  f_1(x)=-\nu\tanh(\nu x).
  \label{f1}
\ee
Substituting into Eq.~(\ref{U1}) yields
\be
  U_1=\nu^2\left[1-2\textrm{sech}^2(\nu x)\right]+\kappa_1^2.
  \label{U1f}
\ee

To find $U_2$ we consider
\be
  U_2=f_2^2+f'_2+\kappa_2^2
  \label{U2}
\ee
and solve
\be
  -w_2''+U_1w_2=\kappa_2^2w_2.
  \label{w2}
\ee
To solve Eq.~(\ref{w2}) we use Eq.~(\ref{U1f}). Now defining $z\equiv\nu x$ and  $\beta^2\equiv\kappa_1^2-\kappa_2^2$ we can then write Eq.~(\ref{w2}) as
\be
  \frac{d^2w_2}{dz^2}+\left(\lambda+2\textrm{sech}^2(z)\right)w_2=0
\ee
where
\be
  \lambda\equiv-\left(1+\frac{\beta^2}{\nu^2}\right).
  \label{l3}
\ee
In general the solution for $w_2$ involves the hypergeometric function, however it is instructive to consider a special case. If  we take $\beta^2/\nu^2=3$, the solution for $w_2$ is $w_2=\cosh^2(\nu x)$, therefore
\be
  f_2(x)=-2\nu\tanh(\nu x).
\ee
Substituting into Eq.~(\ref{U2}) yields
\be
  U_2=2\nu^2\left[2-3\textrm{sech}^2(\nu x)\right]+\kappa_2^2.
  \label{U2f}
\ee

To find $U_3$ we consider
\be
  U_3=f_3^2+f'_3+\kappa_3^2
  \label{U3}
\ee
and solve
\be
  -w_3''+U_2w_3=\kappa_3^2w_3.
  \label{w3}
\ee
To solve Eq.~(\ref{w3}) we use Eq.~(\ref{U2f}). Now defining $z\equiv\nu x$ and define $\gamma^2\equiv\kappa_2^2-\kappa_3^2$ we can then write Eq.~(\ref{w3}) as
\be
  \frac{d^2w_3}{dz^2}+\left(\sigma+6\textrm{sech}^2(z)\right)w_3=0
\ee
where $\sigma\equiv-(4+\gamma^2/\nu^2)$. In general the solution for $w_2$ involves the hypergeometric function, however it is more instructive to consider a special case. If  we take $\gamma^2/\nu^2=5$, the solution for $w_3$ is $w_3=\cosh^3(\nu x)$, therefore
\be
  f_3(x)=-3\nu\tanh(\nu x).
\ee
Substituting into Eq.~(\ref{U3}) yields
\be
  U_3=3\nu^2\left[3-4\textrm{sech}^2(\nu x)\right]+\kappa_3^2.
  \label{U3f}
\ee

Finally to find $U_4$ we consider
\be
  U_4=f_4^2+f'_4+\kappa_4^2=f_4^2+f'_4
  \label{U4}
\ee
and solve
\be
  -w_4''+U_3w_4=\kappa_4^2w_4.
  \label{w4}
\ee
To solve Eq.~(\ref{w4}) we use Eq.~(\ref{U3f}). Now using $z\equiv\nu x$ we can  write Eq.~(\ref{w4}) as
\be
  \frac{d^2w_4}{dz^2}+\left(\rho+12\textrm{sech}^2(z)\right)w_4=0
\ee
where $\rho\equiv-(9+\kappa_3^2/\nu^2)$. In general the solution for $w_4$ involves the hypergeometric function, however it is more instructive to consider a special case. If  we take $\kappa_3^2/\nu^2=7$, the solution for $w_4$ is $w_4=\cosh^4(\nu x)$.

The profile function is defined as
\be
  \phi_0=\alpha\int\frac{dx}{w_N}
\ee
where $N$ is the highest eigenvalue, in this case 4, and $\alpha$ is a normalization constant. Therefore here we can write,
\be
  \phi_0=\frac{\alpha}{3}\tanh(\nu x)\left(\textrm{sech}^2(\nu x)+2\right).
\ee
For significantly high values of $\nu x$ this can be approximated as
\be
  \phi_0\approx\frac{2\alpha}{3}\tanh(\nu x).
  \label{phi1}
\ee
The symmetry breaking potential is then defined from the Bogomolnyi equation
\be
  V(\phi_0)=\frac{1}{2}\psi_0^2(x)\Big{|}_{x(\phi_0)}
\ee
where
\be
  \psi_0(x)=\frac{d\phi_0}{dx}.
\ee
Therefore we can write the symmetry breaking potential as
\bea
  V(\phi_0)&=&\frac{\alpha^2}{2}\textrm{sech}^8(\nu x)\nonumber\\
    &=&\frac{\alpha^2}{2}\left(1-\frac{9\phi_0^2}{4\alpha^2}\right)^4
\eea
where we used Eq.~(\ref{phi1}). As expected, the potential in question is $\phi^8$. However, the conditions that we imposed on the eigenvalues introduced a high level of symmetry in the potential. In order to get a desired potential with two pairs of non-equivalent minima we need to relax the conditions on the eigenvalues.

Now, we consider a more generic case of the potential that we need to reconstruct. In this case we search for solutions to our given potential, Eq.~(\ref{pot}). Such as in the special case, the potential $U_1$ is given by Eq.~(\ref{U1}). Using this potential, we can again solve the differential equation for $w_1(x)$. The solution for $w_1(x)$ is given by $w_1(x)=\cosh(\nu x)$, which then yields $f_1$, Eq.~(\ref{f1}). From Eq.~(\ref{U1f}) and Eq.~(\ref{w2}), we can then solve the differential equation which has a solution
\be
  w_2(x)=P_1^{\sqrt{-\lambda}}(\tanh(\nu x))+Q_1^{\sqrt{-\lambda}}(\tanh(\nu x))
  \label{gen w2}
\ee
where $\lambda$ is the same as in the special case above, Eq.~(\ref{l3}), and $P$ and $Q$ are the Legendre polynomials of the first and second kind, respectively. From the parity condition, we can see that for the Legendre polynomials of the second kind $\sqrt{-\lambda}$ must be an even integer. For the Legendre polynomial of the first kind, the only non-zero terms are for $\sqrt{-\lambda}=0,1$. For $\sqrt{-\lambda}=0$, this does not satisfy the parity condition, hence it is not an allowed solution. For $\sqrt{-\lambda}=1$, we then have the solution
\be
  w_2(x)=-\sqrt{1-\tanh(\nu x)^2}=\pm\textrm{sech}(\nu x)\nonumber.
\ee
This then gives that the function $f_2$ is given by
\be
  f_2=\pm\nu\tanh(\nu x).\nonumber
\ee
This is just $\mp f_1$, hence when we solve for $w_3(x)$ we will just obtain Eq.~(\ref{gen w2}). We will end up going in circles following this value; it is therefore more economic to explore the Legendre polynomial of the second kind.

From Eq.~(\ref{gen w2}) we can find the function $f_2$ to be
\bea
  f_2&=&(2-\sqrt{-\lambda})\nu Q_2^{\sqrt{-\lambda}}(\tanh(\nu x))Q_1^{\sqrt{-\lambda}}(\tanh(\nu x))\nonumber\\
      &&-2\nu\tanh(\nu x).
  \label{gen f2}
\eea
From Eq.~(\ref{gen f2}), we can then find the potential $U_2$ using Eq.~(\ref{U}). We use Eq.~(\ref{gen f2}) to find $U_2$, then use this to find $w_3(x)$. However, there is no closed form solution for $w_3(x)$ and we need to use numerical methods from here out to find the solution for both $w_3(x)$ and $w_4(x)$. Using $w_4(x)$, we can then find the reconstructed potential $V(\phi_0)$ which we show in Fig.~\ref{Recon}.
The specific values that give the characteristic shape in question are $\nu=1$, $\lambda=-4$, $\kappa_2^2-\kappa_3^2=5$ and $\kappa_3^2-\kappa_4^2=\kappa_3^2=3$. For convenience, we reconstructed the potential in two intervals, the first one between $\phi =0$ and $\phi =0.8$TeV, and the second one between $\phi =0.8$TeV and $\phi =\infty$. We then joined the plots into a single plot in Fig.~\ref{Recon}.

\section{Kink solution of the field $\phi$ between the vacua $\phi_1$ and $\phi_2$}

The potential given in Eq.~(\ref{Veps}) contains a kink-like solution interpolating between the vacua $\phi_1$ and $\phi_2$.  Setting  $\varepsilon_0 =0$ in the Bogomolnyi equation we find
\be
\sqrt{\frac{\lambda_3}{8}}x=\int \frac{d\phi}{(\phi^2-\phi_1^2)(\phi^2-\phi_2^2)}.
\ee
We evaluate this integral using the method of partial fractions,
\be
\sqrt{\lambda_3}(\phi_2^2-\phi_1^2)x=\ln \left[\left(\frac{\phi+\phi_1}{\phi-\phi_1}\right)^\frac{1}{\phi_1}\left(\frac{\phi-\phi_2}{\phi+\phi_2}\right)^\frac{1}{\phi_2}\right].
\ee
This gives us $x(\phi)$. We then numerically invert it to get  $\phi(x)$. The result is shown in Fig.~\ref{px}.
\begin{figure}
\includegraphics[width=3in]{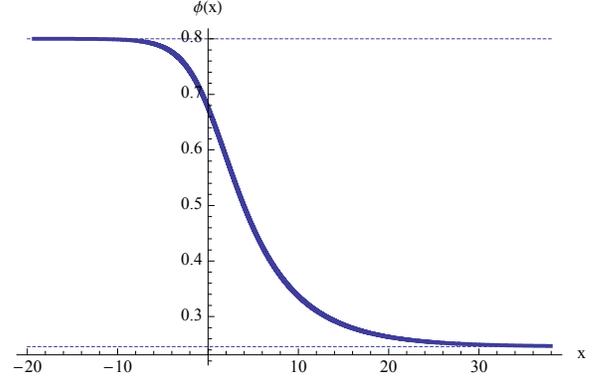}
\caption{A kink-like solution interpolating between the vacua $\phi_1$ and $\phi_2$ for  $\varepsilon_0 =0$.}
\label{px}
\end{figure}

\section{Critical temperature}
\label{ct}

Several effects may change the numerical value of the critical temperature of the phase transition. We first consider the possibility that the two minima do not contain the same number of degrees of freedom. Let us denote the number of degrees of freedom in the two vacua $\phi_1$ and $\phi_2$ by $N_1$ and $N_2$ respectively. The condition for the critical temperature is that the values of the effective potential (\ref{veff}) evaluated at the two vacua $\phi_1$ and $\phi_2$ are the same. From (\ref{veff}), after we add the terms that depend on N, we have

\be \label{veff1}
  V_{eff}(\phi_1,T) = -\left( \varepsilon_0-\frac{cT^2}{\phi_1^3\phi_2^3} \right) \phi_1^5\phi_2^3 -\frac{\pi^2}{90}N_1T^4+{\rm const}.
\ee
\be \label{veff2}
  V_{eff}(\phi_2,T) = -\left( \varepsilon_0-\frac{cT^2}{\phi_1^3\phi_2^3} \right) \phi_1^3\phi_2^5 -\frac{\pi^2}{90}N_2T^4+{\rm const}.
\ee
where $N_i$, $i=1,2$, represent the contribution over the relativistic bosonic, $N_B$, and fermionic, $N_F$, spin states, i.e.
\be
N= N_b + \frac{7}{8}N_F.
\ee
At $T=T_c$, we have $V_{eff}(\phi_1,T_c) =  V_{eff}(\phi_2,T_c)$ which gives a quartic equation for $T_c$
\be \label{eqtc}
(N_1-N_2)T_c^4+c(\phi_2^2-\phi_1^2)T_c^2-\varepsilon_0 \phi_1^3\phi_2^3 (\phi_2^2-\phi_1^2) =0
\ee
We can now solve Eq.~(\ref{eqtc}) for $T_c$ to get
\be
T_c^2 \!= \!\frac{-c (\phi_2^2-\phi_1^2) \! \pm \! \sqrt{c^2(\phi_2^2-\phi_1^2)^2 +\frac{2\pi^2}{45}(N_1-N_2)\varepsilon_0 \phi_1^3\phi_2^3 (\phi_2^2-\phi_1^2) }   }{\frac{\pi^2}{45}(N_1-N_2)}
\ee
We can perform a simple estimate of $\Delta N = N_1 - N_2$ if we assume that in the vacuum $\phi_1$ all the particles are relativistic, while in the vacuum $\phi_2$ the Higgs boson, $W^\pm$ and $Z$ bosons and top quark masses become larger than the temperature of the universe and their contribution should be excluded. Therefore $\Delta N \sim 15$.

Another consequence of the fact that particles not much lighter than the temperature should be decoupled from the plasma is that the constant $c$ defined in Eq.~(\ref{c}) gets modified. If the temperature during the electroweak phase transition is of the order of $100$GeV, then it is reasonable to exclude the Higgs boson, $W^\pm$ and $Z$ bosons and top quark contributions in Eq.~(\ref{c}). This is especially true in the second vacuum $\phi_2$. We first need to express the constant $c$ in terms of physical quantities. For this purpose, we eliminate $\lambda_1$ using Eq.~(\ref{mh}), i.e.
\be
\lambda_1=\frac{1}{3}m_H^2/v^2+\frac{1}{3}\mu^2/v^2+\frac{5}{4}\lambda_2v^2-\frac{7}{6}\lambda_2v^4.
\ee
The constant $c$ now becomes
\begin{eqnarray}
  c=&&\frac{1}{16}\left(3g^2+g'^2+4y_t^2+\frac{1}{96}\mu^2/v^2+ \frac{1}{96}m_H^2/v^2 \nonumber \right. \\ && \left. +\frac{5}{128}\lambda_2v^2-\frac{7}{192}\lambda_3v^4\right) .
\end{eqnarray}
We now exclude the Higgs boson, $W^\pm$ and $Z$ bosons and top quark contributions to get
\be
  c=\frac{1}{16}\left(4y_b^2+\frac{1}{96}\mu^2/v^2+\frac{5}{128}\lambda_2v^2-\frac{7}{192}\lambda_3v^4\right)
\ee
where $y_b$ is the b-quark Yukawa coupling, which we needed to include once the top quark is excluded.
In turn, the critical temperature defined in Eq.~(\ref{tc}) will change. While the numerical value of the critical temperature will change, the qualitative features will remain unchanged.

\end{document}